\documentclass[aip,apl,reprint,unsortedaddress,superscriptaddress]{revtex4-1}

\usepackage{color}
\usepackage{epsfig}
\usepackage{epstopdf}
\usepackage{amsmath}

\usepackage{graphicx}
\usepackage{dcolumn}
\usepackage{bm} 
\usepackage{keyval}
\usepackage{gensymb}
\usepackage{nicefrac}
\usepackage{units}
\usepackage{hyperref}

\begin{document}

\preprint{APS/123-QED}

\title{Chiral anisotropic magnetoresistance of ferromagnetic helices}

\author{Henrik Maurenbrecher}
\affiliation{Institute of Robotics and Intelligent Systems, ETH Zurich, 8092 Zurich, Switzerland}
\affiliation{Department of Materials, ETH Zurich, 8093 Zurich, Switzerland}
\author{Johannes Mendil}%
\affiliation{Department of Materials, ETH Zurich, 8093 Zurich, Switzerland}
\author{George Chatzipirpiridis}%
\affiliation{Institute of Robotics and Intelligent Systems, ETH Zurich, 8092 Zurich, Switzerland}
\author{Michael Mattmann}
\affiliation{Institute of Robotics and Intelligent Systems, ETH Zurich, 8092 Zurich, Switzerland}
\author{Salvador Pan\'{e}}%
\affiliation{Institute of Robotics and Intelligent Systems, ETH Zurich, 8092 Zurich, Switzerland}
\author{Bradley J. Nelson}%
\affiliation{Institute of Robotics and Intelligent Systems, ETH Zurich, 8092 Zurich, Switzerland}
\author{Pietro Gambardella}%
\affiliation{Department of Materials, ETH Zurich, 8093 Zurich, Switzerland}
\date{\today}

\begin{abstract}
We investigate the anisotropic magnetoresistance (AMR) of ferromagnetic CoNi microhelices fabricated by electrodeposition and laser printing. We find that the geometry of the three-dimensional winding determines a characteristic angular and field-dependence of the AMR due to the competition between helical shape anisotropy and external magnetic field. Moreover, we show that there is an additional contribution to the AMR that scales proportionally to the applied current and depends on the helix chirality. We attribute this contribution to the self magnetic field induced by the current, which modifies the orientation of the magnetization relative to the current flow along the helix. Our results underline the interest of three-dimensional curved geometries to tune the AMR and realize tubular magnetoresistive devices.
\end{abstract}

\maketitle

The anisotropic magnetoresistance (AMR) has been intensively studied in the last decades to provide insight into the charge and spin transport properties of magnetic materials\cite{Mcguire1975ieeetm,campbell1982transport} as well as to realize magnetic field sensors\cite{kuijk1975barber,lenz2006magnetic} and recording devices.\cite{tsang1984magnetics,olejnik2017antiferromagnetic} As most of AMR sensors are thin films, this effect is well-known and characterized for planar two-dimensional systems. Recently, advancement in micro- and nanofabrication techniques has opened new pathways for the investigation of the magnetic properties of more complex three-dimensional structures,\cite{streubel2016magnetism,smith2011magnetic,tottori2012magnetic} in which curvilinear or chiral geometries potentially allow for new manifestations of the AMR.

Previous studies have addressed magnetoresistive effects in curved structures such as ferromagnetic nanotubes\cite{ruffer2012magnetic,ruffer2014anisotropic} and rolled-up membranes.\cite{muller2012towards,schumann2012magnetoresistance,streubel2014magnetic}
In these systems, the AMR is mainly determined by the shape-dependent behavior of the magnetization in an external field. In order to further exploit the interplay between structure and magnetotransport, it is appealing to engineer microscale devices with a geometry that allows for additional AMR effects due, e.g., to chiral domain walls\cite{Yin2017} or self-induced current dependent magnetic fields and scattering by nonmagnetic chiral defects.\cite{rikken2001electrical,krstic2002helix} The latter effects are expected to induce nonlinear contributions to the AMR that result in a directional dependence of the resistance on the current flowing in the ferromagnetic structure, similar to the unidirectional spin Hall magnetoresistance reported in ferromagnetic/nonmagnetic bilayers.\cite{avci2015unidirectional,avci2015magnetoresistance,olejnik2015electrical,yasuda2016large} In nonmagnetic conductors, such chiral nonlinear effects have been observed in the ordinary magnetoresistance of bismuth helices,\cite{rikken2001electrical} carbon nanotubes,\cite{krstic2002magneto} and enantiopure molecular crystals.\cite{pop2014electrical}

In this work, we study the AMR in CoNi microhelices of left and right chirality. We find that the helical geometry imprints a characteristic angular dependence on the AMR due to the competition between shape-anisotropy, which favors the direction tangential to the helical path, and external field. This competition results in a different angular dependence of the AMR of the helices compared to planar thin films\cite{Mcguire1975ieeetm,campbell1982transport} and tubular structures.\cite{ruffer2012magnetic,ruffer2014anisotropic,muller2012towards,schumann2012magnetoresistance,streubel2014magnetic} Moreover, we find that the helical geometry gives rise to a solenoidal field that interacts with the magnetization, inducing a unidirectional contribution to the AMR that depends on the helix chirality. In the following, we describe the realization of the CoNi helices and report on the AMR measurements as a function of current and magnetic field strength.

\begin{figure}
	\centering
	\includegraphics[width=0.45\textwidth]{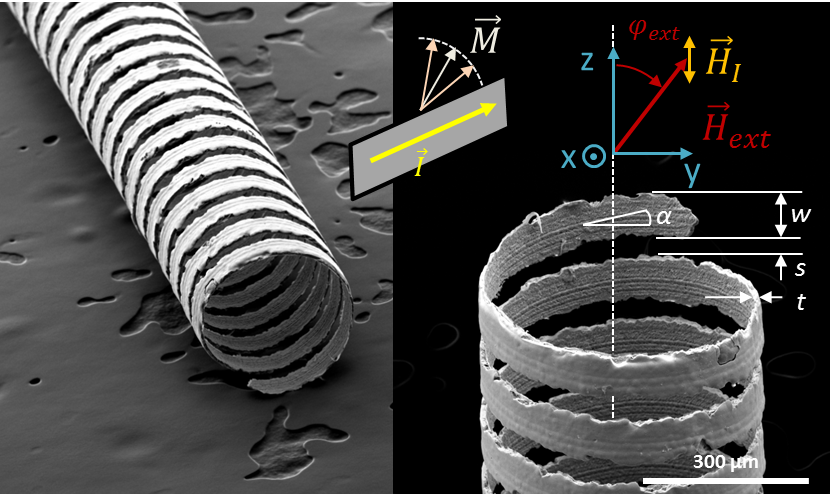}
	\caption{Scanning electron microscopy images of a left-handed helix and coordinates of the reference system.}
	\label{fig:setup}
\vspace{-0.3cm}
\end{figure}

The helices were fabricated by means of electroforming using a Cu mandrel with a diameter of 500\,$\mu$m.\cite{tottori2012magnetic,Chatzipirpiridis2015} The wire was coated with ethyl cyanoacrylate super-glue, onto which the helical pattern was printed by a laser beam that removed the coating while rotating the wire and letting the laser run along it. The pitch angle was controlled by varying the rotational speed of the wire and the linear speed of the laser. The resulting template was then used for the electro-deposition of CoNi into the conductive trenches. Finally, the Cu wire was chemically wet-etched, allowing the helices to be removed from the support. The alloy stoichiometry determined by energy-dispersive x-ray spectroscopy is Co$_{40}$Ni$_{60}$, which is known for its large AMR in bulk specimens.\cite{Mcguire1975ieeetm} The helices are polycrystalline with mixed face-centered cubic, hexagonal-close-packed, and a minority of amorphous structures.\cite{Chatzipirpiridis2015,Ergeneman2011} All samples have length, inner diameter, separation, pitch angle, width, and thickness of $l= 10$~mm, $d = 500$~$\mu$m, $s = 70$~$\mu$m, $\alpha = 9\,\degree$, $w = 50$~$\mu$m, and $t= 8, 20$~$\mu$m, respectively (Fig.~\ref{fig:setup}). We present results for two pairs of samples (four helices in total), which differ from each other only in thickness, with the first pair of helices having a thickness of 8~$\mu$m and the second pair of 20~$\mu$m. We will henceforth refer to these samples according to their thickness, Thin/Thick, as well as their chirality, A/B for right-handedness/left-handedness.

To characterize the AMR, we measured the electrical resistance of the helices using a four contact geometry in the presence of an externally applied magnetic field $\vec{H}_{ext}$. We applied an AC current of frequency $f=10$~Hz  and amplitude $I$ ranging from 1 to 20~mA in order to separate the linear (current-independent) and nonlinear (current-dependent) contributions to the sample's resistance using harmonic signal analysis.\cite{avci2015unidirectional} The total resistance can be expressed as $R = R^{1f}+R^{2f}$, where $R^{1f}$ and $R^{2f}$ are the first and second harmonic components of the resistance obtained by Fourier transformation of the voltage measured between the two ends of an helix. Here, $R^{1f}$ includes the conventional AMR, whereas $R^{2f}$ contains additional resistive effects that are proportional to $I$. As applying a current to a helix gives rise to a solenoid-like magnetic field $\vec{H}_{I}\propto I$, one such effect is the perturbation of the magnetization $\vec{M}$ around its equilibrium orientation induced by the self field, which modulates the AMR of the helix in addition to $\vec{H}_{ext}$. In order to quantify the observed effects, we compare the normed peak-to-peak values of $R^{1f}$ and $R^{2f}$, given by
\begin{equation}\label{eq:DeltaR}
	\frac{\Delta R^{1f}}{R_{min}} = \frac{R^{1f}_{max}-R^{1f}_{min}}{R^{1f}_{min}}, \quad
	\frac{\Delta R^{2f}}{R_{min}} = \frac{R^{2f}_{max}-R^{1f}_{min}}{R^{1f}_{min}},
\end{equation}
where $R^{1f,2f}_{max}$ ($R^{1f,2f}_{min}$) are the maxima (minima) of $R^{1f,2f}$.

\begin{figure}
	\centering
	\includegraphics[width=0.5\textwidth]{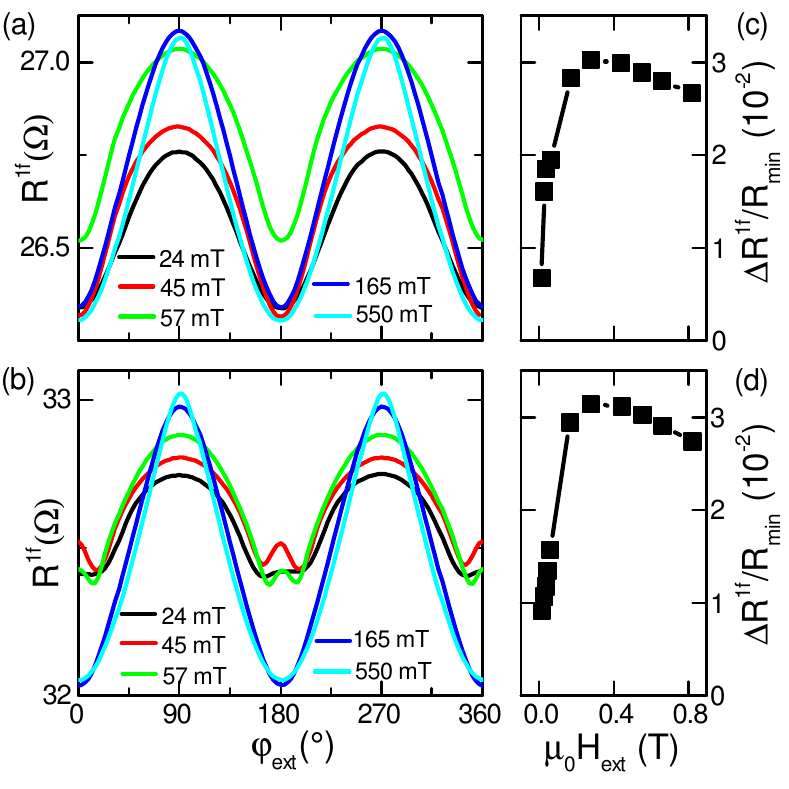}
	\caption{Angular dependence of the AMR on external field. (a) $R^{1f}$ of Thick A and (b) Thin A helices for different strengths of $\vec{H}_{ext}$ measured for a current of amplitude $I=20$~mA. (c,d) Normed peak-to-peak values $\Delta R^{1f} / R^{1f}_{min}$ as a function of applied field.}
	\label{fig:R1f}
\end{figure}

We begin by investigating $R^{1f}$ to see if the AMR has any geometry induced effects. Starting with $\vec{H}_{ext}$ applied along the long axis of the helix ($\varphi_{ext} = 0\,\degree$), the samples are rotated in the $yz$-plane, yielding $R(\varphi_{ext})$. Figure~\ref{fig:R1f} shows $R^{1f}$ plotted as a function of $\varphi_{ext}$ for different field strengths for the samples Thick A (a) and Thin A (b). For both samples, $R^{1f}$ appears to behave similar to the AMR of polycrystalline thin-films, given by
\begin{equation}\label{eq:AMR}
R = R_{\perp} + (R_{\parallel}-R_{\perp})\,(\hat{M} \cdot \hat{I})^{2},
\end{equation}
where $R_{\perp}$ and $R_{\parallel}$ are the resistance for $\vec{M}\perp\vec{I}$ and $\vec{M}\parallel\vec{I}$, respectively. This behavior produces a typical unsaturated bell-shaped curve at low field, which transitions into a $\sin^2 \varphi_{ext}$ lineshape as $\vec{H}_{ext}$ becomes strong enough to force $\vec{M}$ along its orientation. However, an important difference relative to the AMR in thin-films can be seen in the normed amplitude of $R^{1f}$, where $\Delta R^{1f}/R_{min}$ grows linearly until it reaches a maximum at $170\,$mT, and then starts to decline [Fig.~\ref{fig:R1f} (c) and (d)]. This trend can be rationalized by analyzing how $R^{1f}_{max} = R^{1f}(\varphi_{ext} = 90,\,270\degree)$ and $R^{1f}_{min} = R^{1f}(\varphi_{ext} = 0,\,180\degree)$ change as a function of $\vec{H}_{ext}$. The behavior of $R^{1f}_{max}$ can be explained by considering that the preferred orientation of $\vec{M}$ induced by shape anisotropy is the tangential direction, for which $\vec{M}\parallel\vec{I}$. For low fields applied at $\varphi_{ext} = 90,\,270\degree$, $\vec{H}_{ext}$ favors the re-orientation of magnetic domains with $\vec{M}$ parallel to the tangential easy axis, leading to a large $R^{1f}_{max}$. For $\varphi_{ext} = 0,\,180\degree$, on the other hand, the field is aligned with the long axis of the helix and $\vec{M}$ is progressively forced towards $\hat{z}$, leading to a decrease of $R^{1f}_{min}$. As the shape anisotropy in the radial direction is much stronger than along $\hat{z}$, the initial decrease of $R^{1f}_{min}$ is larger than the change of $R^{1f}_{max}$. The difference between $R^{1f}_{max}$ and $R^{1f}_{min}$ leads to an increase of $\Delta R^{1f}$ up to $\vec{H}_{ext} \approx 170\,$ mT. At higher fields, $\vec{M}$ is saturated along $\hat{z}$, but not along the radial direction. Thus, $R^{1f}_{min}$ does not change whereas $R^{1f}_{max}$ slowly decreases with increasing field, leading to a small reduction of $\Delta R^{1f}$.

Another interesting geometric effect can be seen in $R^{1f}$ of the thinner sample [Fig.~\ref{fig:R1f} (b)], where two minima instead of one appear close to $\varphi_{ext} = 0,\,180\degree$, shifted by the pitch angle $\pm\alpha$. This effect can be understood by considering one winding of the helix as the projection onto the $yz-$plane of two mirrored planar elements, each with its own $\vec{M}$. At $\varphi_{ext} = 0,\,180\degree$, $\vec{M}$ is not perpendicular to the current in either of the two elements. 
As $\varphi_{ext}$ moves from $0\degree$ to $\alpha = 9\degree$, $\vec{M}$ in one element increasingly aligns itself to the easy axis, which is also the direction of the current, thus increasing $R^{1f}$ in this element. This increase, however, is negated by the greater rate of change away from the easy axis of $\vec{M}$ in the second element, which results in a net decrease of $R^{1f}$ and a minimum at $\varphi_{ext} = \pm\alpha$. This behavior is only seen in the thin helices, as they have a stronger shape anisotropy compared to the thick ones. Finally, we remark that there is no noticeable influence of the current and helix chirality on $R^{1f}$.

\begin{figure}
	\centering
	\includegraphics[width=0.5\textwidth]{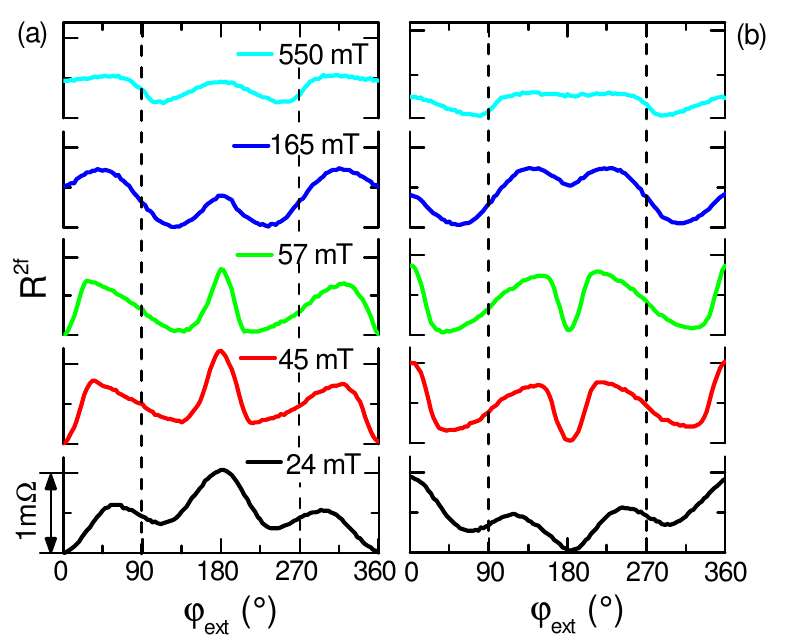}
	\caption{Angular dependence of the current-induced AMR on external field. $R^{2f}$ of (a) Thick A and (b) Thick B helices for different strengths of $\vec{H}_{ext}$ measured for a current of amplitude $I=20$~mA.}
	\label{fig:R2f}
\end{figure}

We now turn our attention to the nonlinear effects in the AMR, captured by $R^{2f}$. Figure~\ref{fig:R2f} shows that there is a distinct current-induced AMR signal that has opposite sign in right- and left-handed helices and has a different angular dependence on $\vec{H}_{ext}$ compared to $R^{1f}$. Whereas the difference in chirality has little or no effect on $R^{1f}$, it clearly produces the mirrored behavior in $R^{2f}$. This is a first indication that $R^{2f}$ arises from a self-induced solenoidal field $\vec{H}_{I}$, which changes sign depending on the winding direction of the helices. The measurements also show that the current-induced effects are largest at low external magnetic fields. This behavior is consistent with a modulation of the magnetization direction caused by $\vec{H}_{I}$, as the relative contribution of this field to the total magnetic field increases for decreasing $\vec{H}_{ext}$.
Indeed, for small $\vec{H}_{ext}$, $|R^{2f}|$ is largest at $\varphi_{ext} = 0,\,180\degree$ as $\vec{H}_{I}$ reinforces or weakens $\vec{H}_{ext}$ in tilting the magnetization perpendicular to the tangential direction. For large $\vec{H}_{ext}$, on the other hand, only the component perpendicular to the external field, $H_{I\bot} = H_{I}\sin\varphi_{ext}$, can change the orientation of $\vec{M}$ in a significant way. As the gradient of $R^{1f}$, $\frac{\partial R^{1f}}{\partial\varphi_{ext}}\propto \sin 2\varphi_{ext}$, describes the linear relationship in the change in resistance for small perturbations around this point, the second harmonic signal is approximately given by $R^{2f}(\varphi_{ext}) \propto H_{I}\sin\varphi_{ext} \sin 2\varphi_{ext}$, as observed in Fig.~\ref{fig:R2f}.

\begin{figure}
	\centering
	\includegraphics[width=0.4\textwidth]{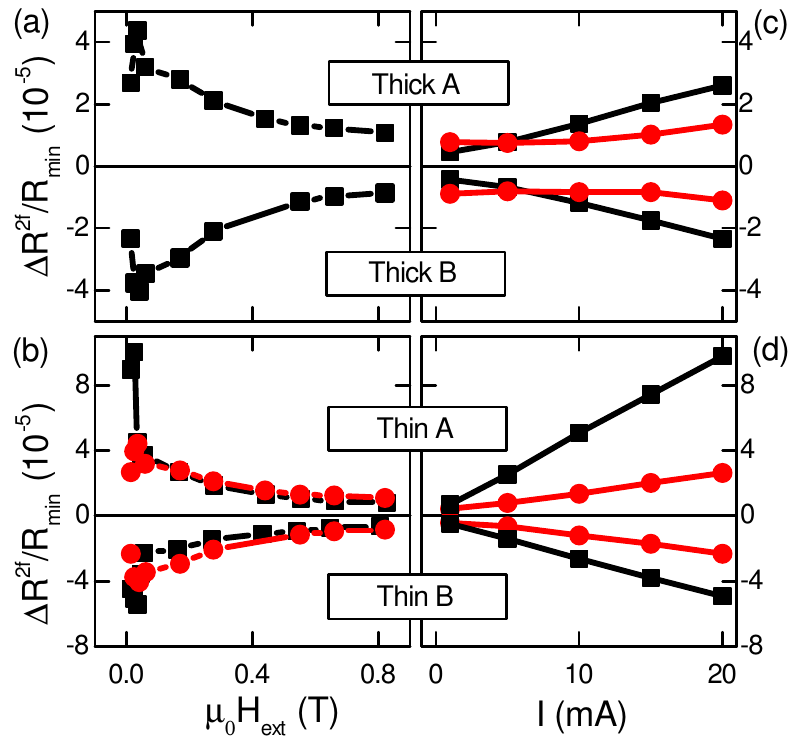}
	\caption{Field dependence of $\Delta R^{2f}/R_{min}$ for (a) Thick and (b) Thin helices of A/B chirality. (c, d) Current dependence of $\Delta R^{2f}/R_{min}$ for the Thick (c) and Thin (d) helices at constant external fields of 13.5\,mT (black) and 550\,mT (red).}
	\label{fig:DeltaR2fandCurrDep}
\end{figure}

Figure~\ref{fig:DeltaR2fandCurrDep} shows the normed peak-to-peak values of $R^{2f}$ of Thick and Thin helices of opposite chirality plotted for different field and current strengths. Besides the strong decrease of $R^{2f}$ with $\vec{H}_{ext}$, we observe that $\Delta R^{2f}/R_{min}$ is linearly proportional to the current amplitude, reaching values of up to 0.01~\% in the Thin helices for $I=20$~mA. The thickness of the helices can also be seen to have an effect on the current-induced AMR, as $\Delta R^{2f}/R_{min}$ of the Thin samples is generally greater than that of the Thick ones for fields below 40\,mT. We attribute this behavior to the thinner samples having a greater shape anisotropy in the tangential direction, which favors the orthogonality between $\vec{H}_{I}$ and $\vec{M}$ at low field, thus augmenting the current-induced deflection of $\vec{M}$ relative to the thicker samples. We note that, according to our interpretation, $\Delta R^{2f}/R_{min}$ is proportional to the absolute current flowing in the helices, rather than to the current density $j$. However, if we normalize the maxima of the curves in Fig.~\ref{fig:DeltaR2fandCurrDep} (a) and (b) by the current density, we find that $\Delta R^{2f}/(R_{min} j) \approx 2 \times 10^{-12}$~m$^{2}$A$^{-1}$ for both Thick and Thin samples, since the shape anisotropy scales approximately with the inverse of the thickness,\cite{aharoni1998demagnetizing} as $j$ does. The difference in the low field magnitude of $R^{2f}$ for samples Thin A and Thin B is not understood and is tentatively assigned to unintentional structural differences between the two samples. Further, we find that $\Delta R^{2f}/R_{min}$ decreases by 14\% in shorter helices with $L= 5$~mm, consistently with the decrease of the axial component of $\vec{H}_{I}$ produced by a finite solenoid of reduced length. For the same reason, we expect that increasing the coil spacing should reduce $R^{2f}$.

\begin{figure}
	\centering
	\includegraphics[width=0.5\textwidth]{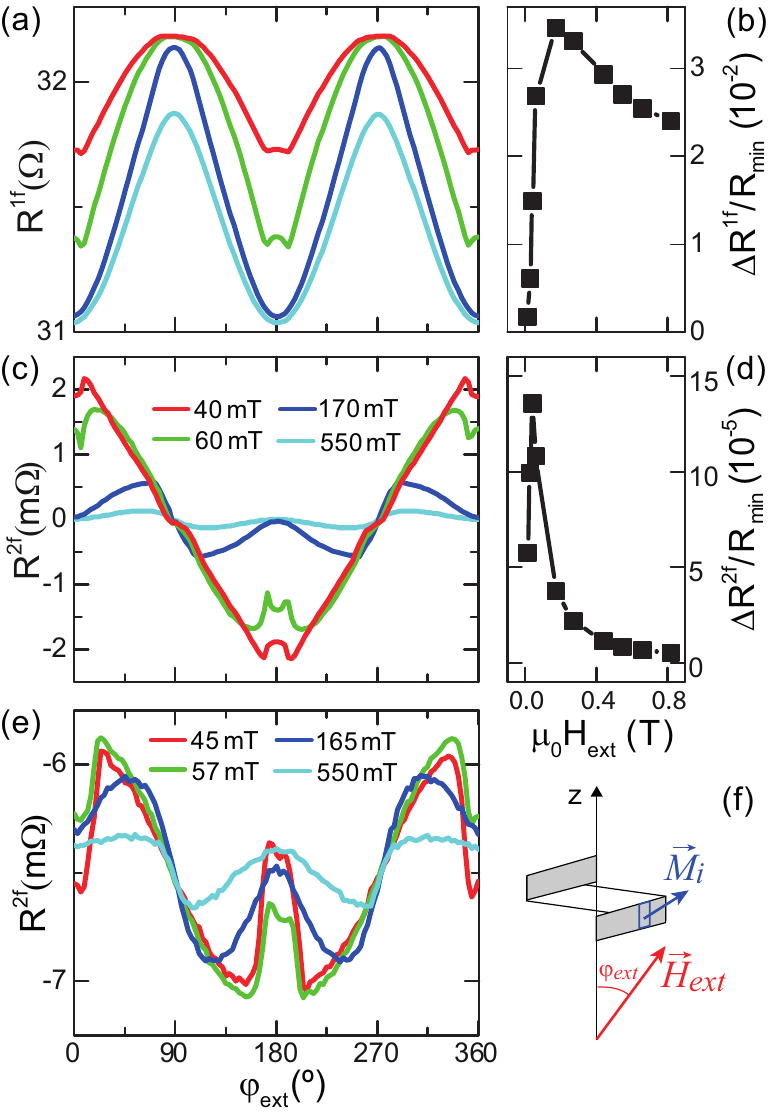}
	\caption{Finite element simulations of (a) the angular dependence of $R^{1f}$ and (b) the field dependence of $\frac{\Delta R^{1f}}{R_{min}}$ for the sample Thin A with an induced field of $\mu_{0}H_{I} = 50$~$\mu$T. (c, d) Simulations of the angular dependence of $R^{2f}$ and field dependence of $\frac{\Delta R^{2f}}{R_{min}}$. (e) Angular dependence of $R^{2f}$ for the sample Thin A measured for a current of amplitude $I=20$~mA. (f) Schematic of the finite element model used in the simulations.}
	\label{fig:Simulation}
\end{figure}

To check our understanding of the AMR, we tested a finite element model in which the helix is discretized into $i = 72$ rectangular prismatic elements, each possessing its own magnetization $\vec{M_{i}}$. We computed the magnetic energy of each element, $E_{mag} = -\vec{M}_{i}\cdot(\vec{H}_{ext}+\vec{H}_{I}-\mathbf{N}\cdot\vec{M}_{i})$, where $\mathbf{N}$ is the demagnetization tensor calculated by Aharoni,\cite{aharoni1998demagnetizing} and minimized it to find the equilibrium direction of $\vec{M}_{i}$. Once the orientation of the magnetization is known, the AMR can easily be determined by summing Eq.~\ref{eq:AMR} over all elements. Thereafter, $R^{1f}$ and $R^{2f}$ are separated by a Fast-Fourier-Transform. The results of our simulations can be seen in Fig.~\ref{fig:Simulation}, where the simulated field dependence of the $R^{1f}$ and $R^{2f}$ for Thin A is shown, assuming that $\vec{H}_{I}$ is that of an ideal infinite solenoid and that $|\vec{M}_{i}|$ is the same for each element and equal to 0.44~MAm$^{-1}$, as measured by vibrating sample magnetometry. Despite these simplifying assumptions, we find that the model reproduces the main features of the experimental data. In particular, the first harmonic characteristics, such as the shifted minima in the thin sample, as well as the declining behavior of $R^{1f}$ at higher fields, see Fig.~\ref{fig:R1f}, can be observed. Furthermore, the amplitude and the main features of the angular dependence of $R^{2f}$ are comparable to $R^{2f}$ measured on sample Thin A, as shown in Fig.~\ref{fig:Simulation} (e). A better fit can possibly be obtained by constructing a micro-magnetic model taking into account the exchange energy, the influence of the curvature, the dipolar interaction between adjacent windings, and the non-solenoidal components of $\vec{H}_{I}$. However, our simple finite-element model already provides adequate insight into the linear and nonlinear AMR of the helices.


In summary, we studied how the geometry of a three-dimensional helical structure manifests itself in the AMR. Our measurements show that the geometry and the current both influence the AMR. The first effect is due to the competition between shape anisotropy and external field that gives rise to a specific angular and field dependence of $R^{1f}$, which is independent of the helix chirality and applied current. The second effect, on the other hand, induces a nonlinear contribution to the AMR that depends on the relative orientation of current and magnetization and changes sign in helices of opposite chirality. The magnitude of this effect, quantified by $\Delta R^{2f}/R_{min}$, scales with $|\vec{H}_{I}|\propto I$ and is largest when $\vec{M} \perp \vec{H}_{I}$ in thin samples with strong shape anisotropy. Such current-induced effects are expected to be general to chiral magnetic conductors and should allow for a better understanding of the magnetotransport properties in these systems. From a practical point of view, ferromagnetic microhelices could find applications as tubular magnetoresistive sensors in microfluidic channels for the in-flow detection of magnetic particles.\cite{moench2011rolled} Besides providing a mechanism to sense the presence and direction of the magnetization, the helices could also be used to generate gradients of magnetic fields to steer the particles along the channels. Further geometry optimizations may lead to enhancements of the observed effects.

We gratefully acknowledge funding by the Swiss National Science Foundation (grant No. 200020-172775).

%

\end{document}